\documentclass[aps,amsmath,amssymb,amsfonts,showpacs,nofootinbib]{revtex4-1}

\usepackage{amsfonts}
\usepackage{amsmath}
\usepackage{color}
\usepackage{graphicx}
\usepackage{dcolumn}
\usepackage{bm}
\usepackage{slashed}

\begin{document}

\def\bb    #1{\hbox{\boldmath${#1}$}}

\title{On the Question of the Point-Particle Nature of the Electron}

\author{
Horace W. Crater$^{1*}$\footnote[0]{${}^*$Email address: hcrater@utsi.edu} 
and Cheuk-Yin Wong$^{2\dagger}$
\footnote[0]{${}^\dagger$Email address: wongc@ornl.gov}
}

\affiliation{${}^1$The University of Tennessee Space Institute,
  Tullahoma, Tennessee 37388 \\
${}^2$Physics Division, Oak Ridge National Laboratory, Oak Ridge, TN 37831 }
\date{\today }

\begin{abstract}
The electron and the positron treated as point particles in the Two
Body Dirac equations of constraint dynamics for QED possess a new and
as yet undiscovered peculiar ${}^1S_0$ bound-state which has a very
large binding energy of about 300 keV, in addition to the usual
${}^1S_0$ positronium state with a binding energy of 6.8 eV. The
production and detection of the peculiar ${}^1S_0$ state provide a
test of the electron point-charge property.  As the peculiar ${}^1S_0$
state lies lower than the usual ${}^1S_0$ state, the peculiar
${}^1S_0$ state can be produced by a two-photon decay of the usual
$^{1}S_{0}$ state.  We estimate the rate of the two-photon decay and
show how it depends on the probability $P_{up}$ of the admixture of
the peculiar component in the predominantly usual ${}^1S_0$
positronium.  The produced peculiar ${}^1S_0$ state in turn
annihilates into two photons with a total c.m. energy of about 723
keV. Thus the signature for this new peculiar ${}^1S_0$ positronium
bound state would be the decay of the usual ${}^1S_0$ state into four
photons, with two energies bunching around 150 and two around 360
keV. Such a four-photon decay of the usual ${}^1S_0$ state will not be
present if the electron and positron are not point particles, or if
the mixing probability $P_{up}$ is very small.

\end{abstract}


\maketitle

\section{Introduction}

Quantum electrodynamics (QED) of light interacting with matter has had
great successes and has been tested to a high degree of accuracy. As
emphasized not the least by Dirac \cite{Dir63}, Feynman \cite{Fey90},
Jaffe \cite{Jaf07}, and many others, it is strange but apparently true
that QED on its own is not mathematically consistent because it is not
asymptotically free at short distances. Infinite charge and mass
renormalizations are required at short distances, ensuring that the
resulting perturbed masses and charges agree with observed
values. From a fundamental point of view, QED raises the puzzling
questions with regard to the meaning of equations involving infinite
constants and mathematically undefined operations of infinite
subtractions.  Dirac \cite{Dir63} expresses the point of view that in
analogy to the Bohr quantum theory, the present day QED formalism will
probably be supplanted ultimately by a formalism that will not embody
infinite charge and mass renormalizations. Thus the\ successes of the
renormalization theory would then be seen to be on the same footing as
the successes of the Bohr orbit theory applied to one-electron
problems. Feynman himself, in a book published one year before he died
\cite{Fey90}, described the QED renormalization scheme in less kind
words as a \textquotedblleft hocus pocus"\ process that has prevented
the proof that QED is mathematically consistent. In fairness to the
memory of Feynman, however, it must be pointed out that when he
learned that there was no Landau pole in QCD, he agreed that unlike
QED, QCD is mathematically self-consistent. \ Perhaps in the hope that
mathematicians may, as in times past with earlier problems, be able to
rectify by more rigorous treatments on the problems of QED, Wightman
\cite{Wig56} and later Jaffe \cite{Jaf66,Jaf07} developed and applied
the methods of axiomatic field theory. The conclusion is that lacking
asymptotic freedom, it is unlikely that QED could \textquotedblleft be
brought fully into the arena of mathematics"\cite{Jaf07}. This still
leaves however Dirac's basic objection related to the appearances of
the infinite subtractions.

In the classical theory of electrodynamics, the self-energy is
infinite for a point particle. Weinberg has pointed out that the
infinite classical self-energy for a point particle should be taken as
a warning of similar problems to come in the nature of infinity
subtractions of the masses and charges for particles in quantum field
theory \cite{wein1}. The point nature of particles is therefore
intimately related to the infinite self-energy and the question of
renormalization. Is it necessary to introduce unknown interactions or
unknown electron structure to remove the ambiguities arising from the
infinities, as was suggested in early classical models by Abraham
\cite{abr} and later Dirac \cite{Dir62}, and the axiomatic field
theory of Wightman \cite{Wig56} and Jaffe \cite{Jaf66,Jaf07}? If so,
what types of interactions and electron structure will be needed to
examine such a problem\footnote{ In contrast, Arnowitz, Deser, and
  Misner have found that only in the limit of a point charge and mass
  do gravitational forces exactly counteract the repulsive
  electrostatic self-forces giving stable and static charged point
  particle configurations \cite{deser}. See also \cite{blinder}.}? Is
it alternatively necessary to bypass the problem of infinite
self-energies by postulating fields which only act on other particles
and only by action-at-a-distance \cite{FW}? One can also avoid
classical point-particle mass and charge singularities by using
Wheeler's geometrodynamical wormhole descriptions of a
\textquotedblleft charge without charge" and a \textquotedblleft mass
without mass" \cite{Whe55,Mis60,Won71}.

Rather than focus on theoretical models related to the point-particle nature, we
look for observable properties of $e^{+}e^{-}$ bound states that may
depend on this point-particle property of the constituents. It is worth
pointing out that the concept of an electron point
charge has been commonly assumed. From the close agreement of
experimental and theoretical electron $ g $-values, an upper limit for
the electron radius of $10^{-17}$ m, may be extracted \cite{Kop95}. It
is reinforced by the absence of a form factor in high-energy electron
scattering measurements with an upper limit of the interaction
distance scale of order $3\times 10^{-19}$ m \cite{L395}, and the
small magnitude of the upper limit of the electron dipole moment, $
d_{e}<8.7\times 10^{-29}$ $e\cdot $cm \cite{Hud11,Bar14}. It is
therefore reasonable to examine the consequences of the point electron
concept and look for related physical observables that may be probed
by experimental measurements.

It is clear that the point charge property has the greatest effects on
the interaction between the electron and the positron at short
distances. In this regard, we note that the magnetic hyperfine
interaction between the electron and the positron in the $S$ state is
given by Eq.\ (5.73) of \cite{Jac62},
\begin{equation}
H_{\mathrm{HFS}}=-\frac{8\pi }{3}{\bb\mu }_{e^{-}}\cdot {\bb\mu }
_{e^{+}}\delta (\mathbf{r}),  \label{1a}
\end{equation}
where the magnetic moments ${\bb \mu}_{e^\pm}$ of $e^{+}$ and $e^{-}$
are related to their spins ${\bb s}_{e^\pm}$ by ${\bb\mu }_{e^{\pm
}}$=$e^{\pm }\mathbf{s}_{e^{\pm }}/mc$. In the spin-singlet
${}^{1}S_{0}$ state for which both $e^{+}$$e^{-}$ and ${\bb s}
_{e^{+}}$$\cdot $${\bb s}_{e^{-}}$ are negative, the above spin-spin
interaction is attractive and singular at short distances and may lead
to observable point-charge effects in $e^{+}e^{-}$ bound states. The
traditional treatment of the above interaction presumes the usual
positronium of radius $1/\alpha m$ and treats the interaction as a
perturbation. It does not touch upon nonperturbative bound-state
effects that can be investigated only by solving the bound-state equation
with the inclusion  of the spin-spin
interaction. However,
the attractive delta-function interaction of Eq. (\ref{1a}) is too
singular to be solved if it is included in the non-relativistic
Schr\"odinger equation. Therefore, the presence of the strongly attractive spin-spin
interaction in the ${}^{1}S_{0}$ state necessitates a proper
nonperturbative relativistic treatment of the two-body bound-state
problem.

The Two Body Dirac Equations (TBDE) of Dirac's constraint dynamics
have been previously tested and found to be a proper formalism to
study relativistic two-body bound states. In QCD, they lead to a good
relativistic description of meson spectroscopy in terms of
quark-antiquark bound states for both light and heavy mesons
\cite{crater2,yin,jim,jpn}. In QED, they yield not only a perturbative
spectra that agree with QED standard results but also distinguish
themselves from other bound-state approaches in their ability to
reproduce these same spectral results by nonperturbative bound-state
methods, both numerically and analytically
\cite{exct,becker,crater2}. They give the singular spin-spin
interaction of Eq.\ (\ref{1a}) in the non-relativistic limit. It is
therefore appropriate to use the TBDE to study point-charge effects in
$e^{+}e^{-}$ bound states.

Using the TBDE equations, we found point-charge effects which appears
as the presence of new positronium bound states, in addition to the
usual positronium states \cite{pec}. Their origin was made clear by
the Schr\"{o}dinger-like equation that comes from the Pauli reduction
of the TBDE (see Eq. (\ref{SLE}) below). In particular the magnetic
spin-spin interaction in the $^{1}S_{0}$ states\footnote{ The spin and
  orbital quantum numbers of the ${}^{1}S_{0}$ state refer to those of
  the $\psi _{+}$ wave function in Eq.\ (\ref{SLE}), a four component
  subset of the full sixteen component spinor.} is indeed very
attractive at short distances, modified by the relativistic structure
of the equations to become less singular in relativistic constraint
dynamics than the delta function in the non-relativistic
approximation\footnote{ For an explicit expression of the relativistic
  spin-spin interaction in the $^{1}S_{0}$ state, see Eq. (21) of
  \cite{pec}.  }.  However,  in the ${}^1S_0$ state, the magnetic interaction exactly cancels
the very repulsive Darwin interaction, resulting in a quasipotential
that behaves as $-\alpha ^{2}/r^{2}$ near the origin, for point
electron and positron. The bound-state equation then admits two
different types of states which we designate as usual and
peculiar. They possess distinctly different properties at short
distances. In particular, the peculiar $^{1}S_{0}$ state, yet to be
observed, has a root-mean-square radius of approximately $1/m$ and a
rest mass approximately $\sqrt{2}m$, in contra-distinction from the
usual $^{1}S_{0}$ state with a root-mean-square radius of $1/(\alpha
m)$ and a mass of approximately $2m-m\alpha ^{2}/4$.

The existence of the usual and peculiar states for the positronium
system poses conceptual and mathematical problems \cite{pec}. If we
keep both sets of states in the same Hilbert space, then each set is
complete by itself, but the two sets of states are not orthogonal to
each other. Our system is thus over-complete. Furthermore, the matrix
element of the scaled invariant mass operator for these states between
states of one type and the states of the other type are not symmetric
and thus the invariant mass operator is not self-adjoint.

With the quasipotential $-\alpha ^{2}/r^{2}$ at short distances for
the $ ^{1}S_{0}$ state as it has been determined by the TBDE constraint
dynamics, both the usual and peculiar states are physically admissible
and there do not appear to be compelling reasons to exclude one of the
two sets as being unphysical. We were therefore motivated to introduce
a ``peculiarity" quantum number $\zeta $, such that $\zeta=+1$ for
usual states that have properties the same as those one usually
encounters in QED, and $\zeta=-1$ for peculiar states. The
introduction of the peculiarity quantum number enlarged the Hilbert
space to contain both usual and peculiar states in a complete set and
made the mass operator self-adjoint.

It should be emphasized that if the electron is not a point particle,
then the peculiar state will not exist. Therefore, an experimental
search of the peculiar states can be used to find out whether the
electron is a point particle or not. As the usual $^{1}S_{0}$ state of
mass $\sim $$2m$ lies above the peculiar state of mass
$\sim $$\sqrt{2}m$, the usual $^{1}S_{0}$ state can decay into the
peculiar $^{1}S_{0}$ state by a $0^{+}$$\rightarrow
$$0^{+}$ transition, with the emission of two photons. Because such a
decay has not yet been observed, it is reasonable to consider the
usual ${}^1S_0$ state to be predominantly a peculiarity $\zeta$=1
state, with a small admixture amplitude $M_{\zeta\zeta'}$ of the
peculiarity $\zeta'$=$-$1 component.  Through its admixture to the
peculiar sector, a state in the usual sector can decay to a state of
lower energy in the peculiar sector. The mixing probability
$P_{up}=|M_{\zeta\zeta'}|^2$, for the usual $^{1}S_{0}$ state to admix
with the peculiar $^{1}S_{0}$ state, can be determined by measuring
the decay rate of (usual $^{1}S_{0}$) $\rightarrow $ (peculiar
$^{1}S_{0}$)$ +2\gamma $.

To assist the determination of the mixing probability, we would like
to evaluate how the the two-photon decay rate from the usual $^1S_0$
state to the peculiar $^1S_0$ state depend on $P_{up}$.  After its
production, the $^1S_0$ peculiar state will promptly annihilate into
two photons. We would like to calculate the rate of annihilation of
the $^1S_0$ peculiar state and identify the signature for the
production of the peculiar state.

In the next section we review the formalism leading to the usual and
peculiar solutions of the TBDE for the $^{1}S_{0}$ state of
positronium. \ In section 3 we obtain an estimate for the decay rate
of the usual $^{1}S_{0} $ state to undergo a meta-stable two-photon
decay into the peculiar $ ^{1}S_{0}$ state. In section 4 we evaluate
the annihilation lifetime of the peculiar $^{1}S_{0}$ state. In
section 5, we present the conclusions and discussions. Relevant
details are given in the Appendix.

\bigskip

\section{Usual and Peculiar Bound State Solutions}

The Two-Body Dirac equations of constraint dynamics give a manifestly
covariant 3D reduction of the Bethe-Salpeter equation for two spin-1/2
particles \cite{cra82}. It provides a route \cite{tod78} around the
Currie-Jordan-Sudarshan \textquotedblleft non-interaction theorem"
\cite{cur} that apparently forbade canonical 4-dimensional treatment
of the relativistic $N$-body problem. \ \ For two particles
interacting through a vector interactions the TBDE are given by
\begin{subequations}
\label{TBDE}
\begin{eqnarray}
\mathcal{S}_{1}\psi &\equiv &\gamma _{51}(\gamma _{1}\cdot (p_{1}-\tilde{A}
_{1})+m_{1})\psi =0, \\
\mathcal{S}_{2}\psi &\equiv &\gamma _{52}(\gamma _{2}\cdot (p_{2}-\tilde{A}
_{2})+m_{2})\psi =0,
\end{eqnarray}
in which $\psi $ is a 16 component spinor. \ The operators are compatible
with 
\end{subequations}
\begin{equation}
\left[ \mathcal{S}_{1},\mathcal{S}_{2}\right] \psi =0,~~\text{implying~}
\tilde{A}_{i}=\tilde{A}_{i}(x_{\bot }).
\end{equation}
Thus the potential is forced to depend on $x_{1}-x_{2}$ only through the
transverse component 
\begin{eqnarray}
x_{\bot }^{\mu } &=&\left( \eta ^{\mu \nu }+\hat{P}^{\mu }\hat{P}^{\nu
}\right) (x_{1}-x_{2})_{\nu },~~x_{\bot }\cdot \hat{P}=0,  \notag \\
P &=&p_{1}+p_{2},  \notag \\
\hat{P} &=&P/\sqrt{-P^{2}}.
\end{eqnarray}
One can further show from these two constraints that 
\begin{equation}
p\cdot P\psi =0.
\end{equation}
Thus, in the center-of-momentum (c.m.) frame where
\begin{equation}
P=\left( w,\mathbf{0}\right) ,
\end{equation}
the relative energy is eliminated ($p\psi =(0,\mathbf{p)}\psi $) and
the relative time does not appear ($x_{\bot }=(0,\mathbf{r)}$). The
compatibility condition, $\ \left[
  \mathcal{S}_{1},\mathcal{S}_{2}\right] \psi =0$, also restricts the
spin dependence\ of $\tilde{A}_{i}^{\mu }$ by determining their
dependence on $\gamma _{1},\gamma _{2}$, \cite{pec}
\begin{equation}
\tilde{A}_{i}^{\mu }=\tilde{A}_{i}^{\mu }(A(r),p_{\perp },\hat{P},w,\gamma
_{1},\gamma _{2}),
\end{equation}
with vector interactions $\tilde{A}_{i}^{\mu }$ that depend on an
invariant $ A(r)$ through the vertex form of $\gamma _{1}\cdot \gamma
_{2}$ $.$ \ $~$ The\ Pauli reduction of the TBDE leads to a covariant
Schr\"{o}dinger-like equation for relative motion with an explicit
spin-dependent potential $\Phi .$ \ In the c.m. system it takes the
form $\ $
\begin{eqnarray}
\mathcal{B}^{2}\psi _{+}&\equiv& \{\mathbf{p}^{2} +\Phi (\mathbf{r,}
m_{1},m_{2},w,\mathbf{\sigma }_{1},\mathbf{\sigma }_{2},\mathbf{L})\}\psi
_{+}  \notag \\
 &=&\biggl \{\mathbf{p}^{2}+2\varepsilon _{w}A-A^{2}+\Phi _{D}+\mathbf{\sigma 
}_{1}\mathbf{\cdot \sigma }_{2}\Phi _{SS}  \notag \\
& &+\mathbf{L\cdot (\sigma }_{1}\mathbf{+\sigma }_{2}\mathbf{)}\Phi _{SO}+(3
\mathbf{\sigma }_{1}\mathbf{\cdot \hat{r}\sigma }_{2}\mathbf{\ \cdot \hat{r}
-\sigma }_{1}\mathbf{\cdot \sigma }_{2})\Phi _{T}  \notag \\
& &+\mathbf{L\cdot (\sigma }_{1}\mathbf{-\sigma }_{2}\mathbf{)}\Phi _{SOD}+i
\mathbf{L\cdot \sigma }_{1}\mathbf{\times \sigma }_{2}\Phi _{SOX}  \notag \\
& & +\mathbf{\sigma }_{1}\mathbf{\cdot \hat{r}\sigma }_{2}\mathbf{\cdot \hat{r}
L\cdot (\sigma }_{1}\mathbf{+\sigma }_{2}\mathbf{)}\Phi _{SOT}\biggr \}\psi
_{+}  \notag \\
& =&b^{2}\psi _{+},  \label{SLE}
\end{eqnarray}
where $\psi _{+}$ is a 4-component spinor subcomponent of 16 component
spinor $\psi $. The quasipotentials $\Phi _{D},\Phi _{SS},\Phi
_{SO},\Phi _{T}$, $\Phi _{SOD},\Phi _{SOX}$, and $\Phi _{SOT}$
correspond to the Darwin, spin-spin, spin-orbit, tensor, spin-orbit
difference, spin-orbit product, and spin-orbit tensor interactions,
respectively. Explicit expressions of these interactions are given in
\cite{pec}. The kinematical variables
\begin{align}
m_{w}& =\frac{m_{1}m_{2}}{w}, \\
\varepsilon _{w}& =\frac{w^{2}-m_{1}^{2}-m_{2}^{2}}{2w},
\end{align}
satisfy 
\begin{equation}
b^{2}=\varepsilon _{w}^{2}-m_{w}^{2}=\frac{1}{4w^{2}}
[w^{4}-2w^{2}(m_{1}^{2}+m_{2}^{2})+\left( m_{1}^{2}-m_{2}^{2}\right) ^{2}],
\end{equation}
which corresponds to the Einstein relation between the energy and
reduced mass for the fictitious particle of relative motion. \ The
effects of an eikonal approximation of all the ladder and cross ladder
diagrams and iterated constraint diagrams are embedded in the
c.m. energy dependencies seen in Eq.\ (\ref{SLE}) \cite{saz97}. In
\cite{pec} a number of properties of the TBDE are listed (see also
\cite{jpn}). Among those that will be of importance here is that the
TBDE provide a covariant 3D framework in which the local potential
approximation consistently fulfills the requirements of gauge
invariance in QED \cite{saz96} and that the Schr\"{o}dinger-like
equation with $\Phi (A=-\alpha /r)$ is responsible for accurate QED
spectral results through order $\alpha ^{4}$ \cite{becker}. \

The QED spectral results can be obtained by solving the radial forms
of Eq.\ (\ref{SLE}), either numerically or analytically.  For
equal-mass systems of $e^+e^-$ in the ${}^{1}S_{0}$ state, the
spin-spin interaction $-3\Phi _{SS}$ is indeed very attractive and
strong at short distances, behaving as $-9/8r^{2}$ (this follows from
Eq. (21) in \cite{pec}). Although this is not as singular as its more
well known non-relativistic delta-function form given in
Eq. (\ref{1a}), by itself it would be regarded as singular since it is
more attractive than $-1/4r^{2}$ and would prevent a nonpertrubative
treatment of this term.  However, for equal-mass systems such as
$e^+e^-$ in the ${}^{1}S_{0}$ state, there is an exact cancellation of
the complete spin-spin term $-3\Phi _{SS}$ with the highly repulsive
Darwin interaction $\Phi _{D}$ \cite{pec}, resulting in a
quasipotential that behaves as $-\alpha ^{2}/r^{2}$ near the origin.
The bound-state equation can thus be treated nonperturbtively.  The
eigenvalue equation for the ${}^1S_0$ state becomes
\begin{equation}
\{-\frac{d^{2}}{dr^{2}}+2\varepsilon _{w}A-A^{2}\}u_{0}=b^{2}u_{0}.
\end{equation}
For a point electron and positron with $A=-\alpha /r$, the above becomes 
\begin{equation}
\{-\frac{d^{2}}{dr^{2}}-\frac{2\varepsilon _{w}\alpha }{r}-\frac{\alpha ^{2}
}{r^{2}}\}u_{0}=b^{2}u_{0}.  \label{exct}
\end{equation}
We can examine the behavior of the wave function at short distances ($
r<<\alpha /2\varepsilon _{w}$), where the above equation behaves as
\begin{equation}
\left\{ -\frac{d^{2}}{dr^{2}}-\frac{\alpha ^{2}}{r^{2}}\right\} u_{0}=0.
\end{equation}
Such a short-distance limit is independent of the chosen gauge \cite{saz96}.
The indicial equation has two types of solutions which will be called usual and
peculiar, 
\begin{align}
u_{+}& \sim r^{\lambda _{+}+1};\text{ }\lambda _{+}=(-1+\sqrt{1-4\alpha ^{2}}
)/2;~~~+~\text{usual}  \notag \\
u_{-}& \sim r^{\lambda _{-}+1};\text{ }\lambda _{-}=(-1-\sqrt{1-4\alpha ^{2}}
)/2~;~~-~\text{peculiar.}
\end{align}
At short distances, the probability is\textit{\ finite} for solutions of
both types, 
\begin{equation}
\psi _{\pm }^{2}d^{3}r=\frac{u_{\pm }^{2}}{r^{2}}r^{2}drd\Omega =u_{\pm
}^{2}drd\Omega =r^{(1\pm \sqrt{1-4a^{2}})}drd\Omega ,
\end{equation}
which indicates that the behaviors of the wave functions of both types
are quantum mechanically acceptable near the origin. If $L\neq 0$ so
that $ L(L+1)-\alpha ^{2}>0$ or if the electron is not a point
particle, then the peculiar solution is not physically
admissible\footnote{ In Schiff's Quantum Mechanics \cite{schiff}, a
  solution similar to the peculiar one discussed here is examined for
  the case of the Klein Gordon equation for the Coulomb system. He
  argues that what we would call the peculiar solution can be
  discarded since the source of the Coulomb attraction is a finite
  sized nucleus of radius $r_{0}$. \ Here the issue is not resolved in
  that way since we are allowing for the possibility that the electron
  and positron are point particles in order to test the consequences
  of that assumption.}.

Both $^{1}S_{0}$ \ bound state solutions can be obtained
analytically. The respective sets of eigenvalues for total invariant
center-of-mass energy (mass) $w_{\pm n}~$($n$ is the principle quantum
number) are \cite{pec} \
\begin{equation}
w_{\pm n}=m\sqrt{2+2/\sqrt{1+{\alpha ^{2}}/{(}n\pm \sqrt{1/4-\alpha ^{2}}-1/2
{)^{2}}}}.  \label{13}
\end{equation}
The eigenvalue of a usual state is obtained by taking the positive
sign of the above equation. Its expansion in powers of $\alpha $ gives
the standard QED perturbative results through order $\alpha ^{4}$
\begin{equation}
w_{+n}=2m-m{\alpha ^{2}}/{4}n^{2}-m\alpha ^{4}/2n^{3}(1-11/32/n)+O(\alpha
^{6}),~n=1,2,3,....
\end{equation}
For the usual ground ($n=1$) state, it gives $w_{+n}\sim 2m-m{\alpha ^{2}}/{4
}$. 

The eigenvalue of the peculiar ($n=1$) ground state is obtained by
taking the negative sign of Eq.\ (\ref{13}). It has a mass
\begin{equation}
w_{-1}=m\sqrt{2+2/\sqrt{1+{\alpha ^{2}}/({1/2}-\sqrt{1/4-\alpha ^{2}}{)^{2}}}
}\sim \sqrt{2}m\sqrt{1+\alpha },
\end{equation}
which represents very tight binding energy on order 300 keV for an $
e^{+}e^{-}$ state. The size of the peculiar ground state is on the
order of a Compton wave length $1/m$ \cite{pec}, much smaller than the
Bohr radius size of the usual positronium ground state. \ Its weak
coupling limit has a total c.m. energy approximately $\sqrt{2}m$
instead of $2m$. \ We point out here that this solution does not have
the usual non-relativistic limit. 

The two $n=1$ wave functions have the respective forms, 
\begin{align}
u_{+}(r)& =c_{+}r^{\lambda _{+}+1}\exp (-\kappa _{+}\varepsilon
_{w_{+}}\alpha r),~\kappa _{+}=\frac{2}{1+\sqrt{1-4\alpha ^{2}}}=\frac{1}{
\lambda _{+}+1},  \notag \\
u_{-}(r)& =c_{-}r^{\lambda _{-}+1}\exp (-\kappa _{-}\varepsilon
_{w_{-}}\alpha r),~\kappa _{-}=\frac{2}{1-\sqrt{1-4\alpha ^{2}}}=\frac{1}{
\lambda _{-}+1}.
\end{align}
Since they are both zero node solutions, they are not orthogonal (although
the inner product is small, $\sim 1/1000$)
\begin{equation}
\langle u_{-}|u_{+}\rangle =\int_{0}^{\infty }dru_{+}(r)u_{-}(r)~\sim
~\alpha ^{3/2}\neq 0.
\end{equation}

How do we reconcile this with expected orthogonality of the
eigenfunctions of a self-adjoint operator corresponding to different
eigenvalues? One can show that the second derivative is not
self-adjoint \cite{pec} in this context!  However, we emphasize the
fact that both sets of usual and peculiar states are quantum
mechanically admissible states.  We admit both types of physical
states into a larger Hilbert space by introducing a new operator
$\hat{\zeta}$ with observable quantum number $\zeta $, which we call
``peculiarity".  This will allow the mass operator to be self-adjoint,
and the set of physically allowed states to become a complete
set.  In particular we let
\begin{align}
\hat{\zeta}\chi _{+}& =\zeta \chi _{+}=+\chi _{+}~~\mathrm{with~eigenvalue~}
\zeta =+1,\text{ usual positronium,}  \notag \\
\hat{\zeta}\chi _{-}& =\zeta \chi _{-}=-\chi _{+}~~\mathrm{with~eigenvalue~}
\zeta =-1,\text{ peculiar positronium,}
\end{align}
with the corresponding spinor wave function $\chi _{\zeta }$ assigned
to the states so that a usual state is represented by the peculiarity
spinor $\chi _{+}$,
\begin{equation}
\chi _{+}=
\begin{pmatrix}
1 \\ 
0
\end{pmatrix}
,
\end{equation}
and a peculiar state is represented by the peculiarity spinor $\chi _{-}$ 
\begin{equation}
\chi _{-}=
\begin{pmatrix}
0 \\ 
1
\end{pmatrix}
.
\end{equation}

With this introduction, a general wave function can be expanded in terms of
the complete set of basis functions $\{u_{+n},u_{-n}\}$ as 
\begin{equation}
\Psi =\sum_{\zeta n}a_{\zeta n}u_{\zeta n}\chi _{\zeta },
\end{equation}
where $n$ represents spin and spatial quantum numbers and $\zeta $ the
peculiarity. \ The variational principle applied to $\mathcal{B}^{2}$
\begin{equation}
\langle \mathcal{B}^{2}\rangle =\frac{\langle \Psi |\mathcal{B}^{2}|\Psi
\rangle }{\langle \Psi |\Psi \rangle },
\end{equation}
would lead to 
\begin{align}
\mathcal{B}^{2}u_{+n}\chi _{+}& =-\kappa _{+n}^{2}u_{+n}\chi _{+}, \\
\mathcal{B}^{2}u_{-n}\chi _{-}& =-\kappa _{-n}^{2}u_{-n}\chi _{-}.  \notag
\end{align}
Thus the introduction of the peculiarity quantum number resolves the
problem of the over-completeness property of the basis states and the
non-self-adjoint property of the mass operator $\mathcal{B}^2$.

For completeness, it is worth pointing out that peculiar states
similar to those described above for $e^{+}e^{-}$ would appear also in
other point-charge equal-mass fermion systems such as $\mu ^{+}\mu
^{-}$.  However for unequal-mass fermion systems, the repulsive Darwin
interaction overwhelms the attractive spin-spin interaction as short
distance.  There would thus be no $\ $attractive $-\alpha ^{2}/r^{2}$
term and no peculiar sector.  The Darwin interaction $\Phi _{D}$ has
dual origins: the retardative effects and the usual zitterbewegung
blurring of the relative coordinate such as appears for the electron in
the hydrogen atom.  The former part has been evaluated for
(spin-zero)-(spin-zero) bound states (e.g. $\pi ^{+}\pi ^{-}$) and
shown to be repulsive \cite{cra84}.  The latter part is also
repulsive but is absent for (spin-zero)-(spin-zero) bound states.

\section{Production of the Peculiar ${}^1S_0$ State}

If the peculiar sector and the usual section are disconnected, the
peculiarity quantum number is strictly conserved and states of one
sector will not make transitions to the other sector. There would be
no way to produce the peculiar states from the usual states. The usual
ground $ ^{1}S_{0}$ state would only undergo the usual two photon
annihilation in about $10^{-10}$ sec as shown in Fig.\ 1.

\begin{figure*}[h]
\centering \includegraphics[scale=0.45]{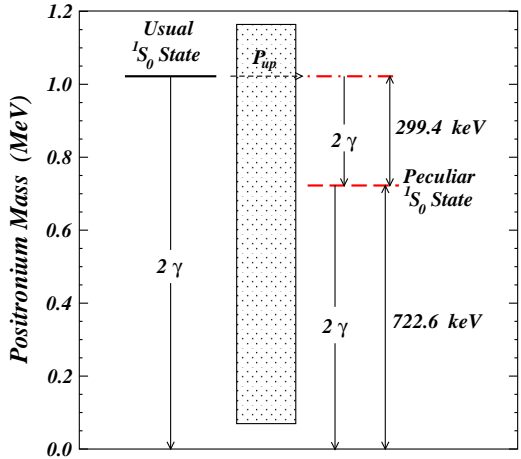}
\caption{(Color online) Schematic energy level diagram for the
  production and detection of the peculiar ${}^1S_0$ state. The usual
  positronium normally decays by annihilation with the emission of 2
  photons.  Through its admixture to the peculiar sector with an an
  admixture probability $P_{up}$, the usual ${}^21S_0$ state can decay
  to the peculiar ${}^1S_0$ state at $w $$\sim$$\sqrt{2}m$ with the
  emission of two photons.  The peculiar ${}^1S_0$ state will
  subsequently annihilate into two additional photons.}
\end{figure*}

We envisage however that while these two sectors are distinct, the
peculiar quantum number may not be conserved for the full Hamiltonian,
and a physical state is an admixture of a usual state and a peculiar state.
The physical  ${}^1S_0$ state, that is predominantly a  usual  $\zeta$=1 state, may be presumed to be $\sqrt{1-|M_{\zeta \zeta'}|^2}\chi_+ + M_{\zeta \zeta'} \chi_-$
with an  mixing
probability $P_{up}$=$|M_{\zeta \zeta ^{\prime }}|^{2}$  
in the peculiar sector.   Through its admixture to the peculiar sector, a state in
the usual sector can decay to a state  in the peculiar
sector with a  lower energy.  In that case the higher (predominantly) usual $(1{}^{1}S_{0})$ state located
at $ w_{+1}\sim 2m-m{\alpha ^{2}}/{4}$ can undergo a meta-stable
$0^{+}$$ \rightarrow 0^{+}$ decay into the lower (predominantly) peculiar
$(1{}^{1}S_{0})$ state located at $ w_{-1}\sim \sqrt{2}m$ by emitting
two photons, as shown in Fig.\ 1. The subsequent annihilation of the
peculiar $1^{1}S_{0}$ state will result in two additional photons for
a total of four photons, with two energies bunching around 150 and two
around 360 keV. The signals of 4 photon decays of definite energies
thus constitute the signature for the peculiar state.  The rate of the
peculiar state production allows the determination of the mixing
probability $P_{up}$.

For brevity of notations, we shall abbreviate the usual ground $
(1{}^{1}S_{0})$ state by $1S_{u}$ and the peculiar ground
$(1{}^{1}S_{0})$ state by $1S_{p}$, where the $2s+1$ superscript index
and the $J$ subscript index are made implicit except when they are
needed to resolve ambiguities.  Having assumed the mixing
probability $P_{up}$, we proceed to determine how the rate of
production of the peculiar $1S_{p}$ state through $1S_{u}\rightarrow
1S_{p}+2\gamma $ depends on $P_{up}$.

With the decaying usual ground spin-singlet state $1S_{u}$ initially
at rest, the decay of the usual $1S_{u}$ state is a three-body decay.
The
produced peculiar $1S_{p}$ state would experience some recoil from the
metastable decay and would have a differential transition rate of
\begin{align}
& d\Gamma ({}1S_{u}\rightarrow {}1S_{p}+2\gamma )  \notag \\
& =2\pi \left\vert T_{{}1S_{u},{}1S_{p}+2\gamma }\right\vert
d^{3}k_{1}d^{3}k_{2}d^{3}p_{1S_{p}}\delta (E_{{}1S_{u}}-E_{{}1S_{p}}-\hbar
\omega _{1}-\hbar \omega _{2})\delta (\mathbf{0}-\mathbf{p}_{{}1S_{p}}-
\mathbf{\ \ k}_{1}-\mathbf{k}_{2}),  \label{rt}
\end{align}
where $T_{{}1S_{u},{}1S_{p}+2\gamma }$ is the transition amplitude for
the process ${}1S_{u}\rightarrow {}1S_{p}+2\gamma $ with the emission
of photons characterized by $(\mathbf{k}_{1},\alpha _{1})$ and
$(\mathbf{k}_{2},\alpha _{2})$. We work in the c.m. system of the
initial positronium atom and use
\begin{equation}
E_{{}1S_{p}}=\sqrt{\left( \sqrt{2}m\right) ^{2}+\mathbf{p}_{{}1S_{p}}^{2}}.
\end{equation}
Performing the $d^{3}p_{{}1S_{p}}$ integral gives
\begin{equation}
{d^{2}\Gamma ({}}1S_{u}\rightarrow {}1S_{p}+2\gamma {)}=2\pi \lvert
T_{{}1S_{u},{}1S_{p}+2\gamma }(\omega _{1},\omega _{2})\lvert
^{2}d^{3}k_{1}d^{3}k_{2}\delta (2m-E_{{}1S_{p}}-\omega _{1}-\omega _{2}),
\label{rate}
\end{equation}
in which 
\begin{align}
E_{{}1S_{p}}& =\sqrt{2m^{2}+\omega _{1}^{2}+\omega _{2}^{2}+2\omega
_{1}\omega _{2}\cos \theta _{12}},  \notag \\
\cos \theta _{12}& =\cos \theta _{1}\cos \theta _{2}+\sin \theta _{1}\sin
\theta _{2}\cos (\phi _{1}-\phi _{2}).
\end{align}

We use perturbation theory to evaluate the transition matrix element $
T_{1S_{u}\rightarrow 1S_{p}+2\gamma }$. The perturbative interaction
leading to the transition is determined by considering
Eqs.\ (\ref{TBDE}) and (\ref {SLE}) and adding an external quantized
space-like photon vector potentials $ A_{\bot }(x_{i})$ orthogonal to
$P,$ in order to define an effective decay Hamiltonian for our
two-body system. The minimal substitution in Eq. (\ref {TBDE}) leads
to
\begin{align}
\gamma _{51}(\gamma _{1}\cdot (p_{1}-\tilde{A}_{1}-eA_{\bot
}(x_{1}))+m_{1})\psi & =0,  \notag \\
\gamma _{52}(\gamma _{2}\cdot (p_{2}-\tilde{A}_{2}+eA_{\bot
}(x_{2}))+m_{2})\psi & =0.
\end{align}
The problem of compatibility of the two separate Dirac equations,
without the presence of an external potential, has been solved and its
Pauli reduction leads to Eq.\ (\ref{SLE}). We can determine the form
that Eq.\ (\ref{SLE}) would subsequently take on if the space-like
parts of the constituent momenta are modified by minimal
substitutions.

In order to set up the decay Hamiltonian in the relativistic case of a
two body system with opposite charges we note that the manifestly
invariant form for Eq.\ (\ref{SLE}) is\footnote{ The four vector forms
  of the spin and orbital angular momenta are $\sigma_{i}^{\mu
  }=\varepsilon ^{\mu \nu \kappa \lambda }\sigma _{i\nu \kappa }\hat{
    P}_{\lambda },~L^{\mu }=\varepsilon ^{\mu \nu \kappa \lambda
  }x_{\bot \nu }p_{\kappa }\hat{P}_{\lambda }$.}
\begin{equation}
\left[ p_{\bot }^{2}-b^{2}(w)+\Phi (x_{\bot }\mathbf{,}w\mathbf{,}\sigma
_{1},\sigma _{2},L)\right] \psi _{+}.
\end{equation}
We use the constituent c.m. energies $\varepsilon _{1}/w=\varepsilon
_{2}/w=1/2$. With $m_{1}=m_{2}\equiv m,$ $b^{2}(w)=w^{2}/4-m^{2},$ $P\cdot
p\psi _{+}=0,$ we have 
\begin{align}
\left( p_{\bot }^{2}-b^{2}(w)\right) \psi _{+}& =\left(
p^{2}-b^{2}(w)\right) \psi _{+}=\frac{1}{2}\left( \left( \frac{P}{2}
+p\right) ^{2}+m_{1}^{2}\right) +\frac{1}{2}\left( \left( \frac{P}{2}
-p\right) ^{2}+m_{2}^{2}\right) \psi _{+}  \notag \\
& =\left( \frac{1}{2}\left( p_{1}^{2}+m_{1}^{2}\right) +\frac{1}{2}\left(
p_{2}^{2}+m_{2}^{2}\right) \right) \psi _{+}.  \notag \\
&
\end{align}
We extend this minimally by 
\begin{align}
p_{1}& \rightarrow p_{1}-eA_{\bot }(x_{1}),  \notag \\
p_{2}& \rightarrow p_{2}+eA_{\bot }(x_{2}).
\end{align}
This leads to 
\begin{align}
p_{\bot }^{2}-b^{2}(w)& \rightarrow p_{\bot }^{2}-b^{2}(w)  \notag \\
& -\frac{1}{2}\left[ e\left( p_{\bot }\cdot A_{\bot }(x_{1})+A_{\bot
}(x_{1})\cdot p_{\bot }\right) +e^{2}A_{\bot }^{2}(x_{1})\right]  \notag \\
& -\frac{1}{2}\left[ e\left( p_{\bot }\cdot A_{\bot }(x_{2})+A_{\bot
}(x_{2})\cdot p_{\bot }\right) +e^{2}A_{\bot }^{2}(x_{2})\right] .  \notag \\
&
\end{align}
In the c.m. system ($A_{\bot }=(0,\mathbf{A)}$) then
\begin{align}
\mathbf{p}^{2}-b^{2}(w)+\Phi & \rightarrow \mathbf{p}^{2}-b^{2}(w)+\Phi 
\notag \\
& -\frac{1}{2}\left[ e\left( \mathbf{p\cdot A(x}_{1},t)+\mathbf{A(x}_{1},t
\mathbf{)\cdot p}\right) +e^{2}\mathbf{A}^{2}\mathbf{(x}_{1}\mathbf{)}\right]
\\
& -\frac{1}{2}\left[ e\left( \mathbf{p\cdot A(x}_{2},t)+\mathbf{A(x}_{1},t
\mathbf{)\cdot p}\right) +e^{2}\mathbf{A}^{2}\mathbf{(x}_{2}\mathbf{)}\right]
.  \notag
\end{align}
We define
\begin{align}
H& =\frac{1}{2\mu }\mathbf{p}^{2}+\frac{1}{2\mu }\Phi  \notag \\
& -\frac{1}{4\mu }\left[ e\left( \mathbf{p\cdot A(x}_{1},t)+\mathbf{A(x}
_{1},t\mathbf{)\cdot p}\right) +e^{2}\mathbf{A}^{2}\mathbf{(x}_{1}\mathbf{)}
\right]  \notag \\
& -\frac{1}{4\mu }\left[ e\left( \mathbf{p\cdot A(x}_{2},t)+\mathbf{A(x}
_{1},t\mathbf{)\cdot p}\right) +e^{2}\mathbf{A}^{2}\mathbf{(x}_{2}\mathbf{)}
\right]  \notag \\
H& =H_{0}+H_{int},  \notag \\
H_{int}& =-\frac{1}{4\mu }\left[ e\left( \mathbf{p\cdot A(x}_{1},t)+\mathbf{
A(x}_{1},t\mathbf{)\cdot p}\right) +e^{2}\mathbf{A}^{2}\mathbf{(x}_{1}
\mathbf{)}\right]  \notag \\
& -\frac{1}{4\mu }\left[ e\left( \mathbf{p\cdot A(x}_{2},t)+\mathbf{A(x}
_{1},t\mathbf{)\cdot p}\right) +e^{2}\mathbf{A}^{2}\mathbf{(x}_{2}\mathbf{)}
\right] .
\end{align}
Our desired matrix element is 
\begin{equation}
T_{1S_{u},1S_{p}+2\gamma }=\langle 1S_{p}\gamma \gamma \lvert H_{int}\lvert
1S_{u}\rangle M_{\zeta \zeta ^{\prime }},
\end{equation}
where $M_{\zeta \zeta ^{\prime }}$ is related to the mixing probability $
P_{up}$ by $P_{up}=|M_{\zeta \zeta ^{\prime }}|^2$. In Appendix A we find in
the dipole approximation that 
\begin{equation}
T_{1S_{u},1S_{p}+2\gamma }\simeq \bigg(\frac{e^{2}}{m}\frac{1}{2(2\pi )^{3} 
\sqrt{\omega _{1}\omega _{2}}}\bigg)\bigg[2^{5/4}\alpha ^{3/2}\bigg]M_{\zeta
\zeta ^{\prime }},
\end{equation}
and so 
\begin{equation}
\lvert T_{1S_{u},1S_{p}+2\gamma }(\omega _{1},\omega _{2})\lvert ^{2}\sim
P_{up}\left\vert \frac{e^{2}}{m}\frac{1}{2(2\pi )^{3}\sqrt{\omega _{1}\omega
_{2}}}2^{5/4}\alpha ^{3/2}\right\vert ^{2}.
\end{equation}
Appendix A shows that the above transition amplitude leads to the
lifetime for the transition of $1S_{u}\rightarrow 1S_{p}+2\gamma $ as
\begin{equation}
\tau ({1S_{u}~\rightarrow ~1S_{p}+2\gamma })\sim \frac{\tau
_{1S_{u}~\rightarrow ~2\gamma ~}}{0.152P_{up}}=\frac{8.2\times 10^{-10}}{
P_{up}}~\text{sec},
\end{equation}
corresponding to a branching ratio of 
\begin{equation}
P(1S_{u}\rightarrow 2\gamma ):P({}1S_{u}\rightarrow {1}S_{p}+2\gamma
)=1~:~0.152~P_{up}.
\label{41}
\end{equation}

\section{Annihilation of the Peculiar $^1S_0$ State into Two Photons}

After the peculiar $1S_{p}$ state is produced, it will subsequently
annihilate into two photons with an energy of approximately $360$ keV
each.  In \cite{decay} we presented previously a formalism for
positronium annihilation, especially suited for relativistic wave
functions that have mild singularities at the origin as occurs with
our usual and peculiar wave functions given in Eq. (\ref{upsn}). \ The
formula below for the decay amplitude involves a radial integral over
the wave function. \ The Yukawa-like form containing the lepton mass
$m~$arises from the folding into the the amplitude of the lepton
exchange that appears in the annihilation Feynman diagram. \ This
amplitude gives the leading order correct result $ \Gamma =m\alpha
^{5}/2$ for the usual ground state positronium decay amplitude,
\begin{equation}
\mathcal{F=}\frac{e^{2}}{m(m+w/2)}\sqrt{2}(2\pi )^{3/2}\int_{0}^{\infty
}drr^{2}j_{1}(wr/2)\left[ \left( \frac{\exp (-mr)}{r}\right) ^{\prime }\left[
(w/2+m)^{2}\psi (r)-\frac{\psi ^{\prime }(r)}{r}\right] \right] .
\end{equation}
Using
\begin{align}
\psi (r)& =\frac{\left( m\sqrt{2}\right) ^{3/2}}{\sqrt{4\pi (1-\sqrt{
1-4\alpha ^{2}})!}}(\sqrt{2}rm)^{-(1-\sqrt{(1-4\alpha ^{2}})/2}\exp (-mr/
\sqrt{2}),  \notag \\
\frac{\psi ^{\prime }(r)}{r}& =\psi (r)(-m/\sqrt{2}-(1-\sqrt{(1-4\alpha ^{2}}
)/2r),  \notag \\
\left( \frac{\exp (-mr)}{r}\right) ^{\prime }& =\frac{\exp (-mr)}{r}(-m-1/r),
\end{align}
leads to
\begin{align}
\mathcal{F}& =\mathcal{-}\frac{e^{2}}{m(m+m/\sqrt{2})}\sqrt{2}(2\pi )^{3/2}
\frac{\left( m\sqrt{2}\right) ^{3/2}}{\sqrt{4\pi (1-\sqrt{1-4\alpha ^{2}})!}}
\int_{0}^{\infty }dr\left( \frac{2\sin (mr/\sqrt{2})}{m^{2}r^{2}}-\frac{
\sqrt{2}\cos (mr/\sqrt{2})}{mr}\right)   \notag \\
& (\sqrt{2}rm)^{-(1-\sqrt{(1-4\alpha ^{2}})/2}\exp (-(m+m/\sqrt{2})r)(mr+1)
\left[ (m/\sqrt{2}+m)^{2}+m/r\sqrt{2}+(1-\sqrt{(1-4\alpha ^{2}})/2r^{2}
\right] .  \notag \\
&
\end{align}
We let $x=mr/\sqrt{2}$, and then we have 
\begin{align}
\mathcal{F}& =\mathcal{-}\frac{e^{2}}{(m+m/\sqrt{2})}2(2\pi )^{3/2}\frac{
\left( m\sqrt{2}\right) ^{3/2}}{\sqrt{4\pi (1-\sqrt{1-4\alpha ^{2}})!}}
\int_{0}^{\infty }dx\left( \frac{\sin (x)}{x^{2}}-\frac{\cos (x)}{x}\right) 
\notag \\
& \times (2x)^{-(1-\sqrt{(1-4\alpha ^{2}})/2}\exp (-(\sqrt{2}+1)x)(\sqrt{2}
x+1)\left[ (1/\sqrt{2}+1)^{2}+1/2x+(1-\sqrt{(1-4\alpha ^{2}})/4x^{2}\right] .
\end{align}
To do the integral, we let $x={y}/(1-y)$ and our integral becomes 
\begin{align}
I& =\int_{0}^{1}\frac{dy}{(1-y)^{2}}\left( \frac{\sin (x)}{x^{2}}-\frac{\cos
(x)}{x}\right)  \\
& \times (2x)^{-(1-\sqrt{(1-4\alpha ^{2}})/2}\exp (-(\sqrt{2}+1)x)(\sqrt{2}
x+1)\left[ (1/\sqrt{2}+1)^{2}+1/2x+(1-\sqrt{(1-4\alpha ^{2}})/4x^{2}\right]
\sim 0.416.  \notag
\end{align}
\ Then with 
\begin{equation}
\frac{e^{2}}{(m+m/\sqrt{2})}2(2\pi )^{3/2}\frac{\left( m\sqrt{2}\right)
^{3/2}}{\sqrt{4\pi (1-\sqrt{1-4\alpha ^{2}})!}}\sim 110\alpha m^{1/2},
\end{equation}
we obtain 
\begin{equation}
\mathcal{F\sim }46\alpha m^{1/2},
\end{equation}
and so the annihilation rate is 
\begin{equation}
\Gamma \sim 2100m\alpha ^{2}>>\frac{m\alpha ^{5}}{2},
\end{equation}
much larger than the usual positronium annihilation rate. The physical
reason for this is the significantly smaller size of the peculiar
positronium bound state compared with the usual state. This leads to
an estimated lifetime of the order of
\begin{equation}
\tau _{{}1S_{p}\rightarrow 2\gamma }\sim \frac{1}{2.1\times 10^{3}m\alpha
^{2}}\sim 10^{-21}\text{sec.}
\label{50}
\end{equation}
This implies that we would see $4\gamma $ in two sequential decays
from the usual ${}^{1}S_{0}$ positronium state as the signature of the
production and the decay of the peculiar ${}^{1}S_{0}$ state.  It is
likely that the two annihilation photons would have energy ranges that
would be distinct from the energy range of the meta-stable decay
photons. The occurrence of the peculiar state will be signaled by a
total of four photons, with two photons bunching at $150$ keV and two
at $360$ keV.  Thus, by considerations of Eqs.\ (\ref{41}) and
(\ref{50}), a measurement of the decay rate of the usual ${}^{1}S_{0}$
positronium state into four photons will allow the determination of
$P_{up}$ and the position of the peculiar ${}^{1}S_{0}$ state, if the
electron and positron are point particles.  Failure to find the
peculiar state at the predicted energy would imply that either
electron and positron are not point-like or $P_{up}$ is too small. The
absence of a point-like nature can arise from the electron having a
structure or to other unknown interactions leading to overall less
attractive interaction potentials that do not give quantum
mechanically acceptable double roots of the leading short distance
behavior. It is unlikely that these would result from QED higher-order
corrections beyond $-\alpha ^{2}/r^{2}$ due to the small value of
$\alpha .$

\bigskip

\section{Discussions and Conclusions}

\bigskip

Is the electron a point particle? To answer such a question, we have
examined the consequences of such a property in bound states of an
electron and a positron. We note that in the relativistic treatment of
the electron and the positron as point particles in QED using the
Two-Body Dirac equations in constraint dynamics, the magnetic
interaction between $e^{+}$ and $e^{-}$ in the ${}^1S_0$ state is very
large and attractive at short distance and cancels the large short
distance repulsion arising from the Darwin interaction. As a
consequence, the interaction at short distances behaves as $-\alpha
^{2}/r^{2}$ and admits two physically allowed solutions. There is a
peculiar ${}^{1}S_{0}$ bound-state solution that has a very large
binding energy of about $300$ keV, in addition to the usual
positronium solution with a binding energy of $6.8$ eV.

We propose a search for the existence of this peculiar $1^{1}S_{0}$
state by looking for a four-photon decay of the usual positronium
$1{}^{1}S_{0}$ state. Specifically, we envisage that the peculiar
sector may be admixed with the usual sector with a mixing  probability
$P_{up}$, yet to be determined.  Subsequent decay of the state to the
lower peculiar state at $\sqrt{2}m$ and the prompt annihilation of the
peculiar state will result in four photons, with two energies bunching
at $150$ keV and two at $360$ keV. We estimate that the usual ground
singlet state $1S_{u}$ state can undergo a meta-stable two photon
decay with a branching ratio of about $0.152P_{up}$ compared to the
dominant annihilation channel.

If the peculiar positronium ground $1S_{p}$ state is found, it would
support the idea that the electron and positron are indeed point
particles. If the peculiar ground singlet state is not found this
would indicate that the electron and positron are not point particles
or would alternatively set limits on the magnitude of the mixing
probability $P_{up}$.

It is anti-intuitive that the peculiar ground state has a binding
energy that does not correspond to the usual non-relativistic limit of
order $\alpha ^{2}$ binding energy. Instead, its binding energy of
about $300$ keV is huge on atomic scales. There is some historical
precedent for such anti-intuitive behavior and that was the existence
of negative energy solutions of the Dirac equation for the hydrogen
atom, which clearly are not physically meaningful in the
non-relativistic approximation. \ Of course, since then with QFT being
based only on positive energy particles \cite {Wig56}, the hole model
and negative energy states were discarded. \ Nevertheless, the
signature of the 4 photon decay of the usual positronium to the
peculiar ground state and its subsequent annihilation would be
striking and give strong direct evidence of the point-like nature of
the electron and positron.

On the other hand, the failure to find the peculiar state may provide
an experimental limit on the point nature of the particles and may
stimulate the search for a description of the structure of the
electron. A proper description of an electron structure may help
resolve the problem of infinite subtractions and infinite
renormalization in QED because these large quantities are limited by
the length scale of the electron structure.

\vspace*{0.5cm} \centerline {\bf Acknowledgment} \vskip.5cm

The authors would like to thank Drs. I-Yang Lee and Paul Vetter for
helpful discussions. The research was sponsored in part by the Office
of Nuclear Physics, U.S. Department of Energy.

\appendix

\setcounter{equation}{0} \ \renewcommand{\theequation}{\Alph{section}.
\arabic{equation}}

\section{Details on the $1S_{u}\rightarrow 1S_{p}+2\gamma $   Decay of the Usual Positronium Ground
State}

\bigskip

\ \noindent We examine the transition of the usual $1^{1}S_{0}$ ground
state of positronium (designated by $~{}1S_{u})$ into the peculiar
$1^{1}S_{0}$ ground state (designated by ${}1S_{p}$) by emitting two
photons. The Golden Rule in this case takes the form \cite{wein}

\begin{equation}
d^{3}w=2\pi \lvert T_{fi}\lvert
^{2}d^{3}k_{1}d^{3}k_{2}d^{3}p_{1S_{p}}\delta
(E_{{}1S_{u}}-E_{{}1S_{p}}-\hbar \omega _{1}-\hbar \omega _{2})\delta (
\mathbf{0}-\mathbf{p}_{{}1S_{p}}-\mathbf{k}_{1}-\mathbf{k}_{2})
\end{equation}

\noindent for the emission of photons characterized by $(\mathbf{k}
_{1},\alpha _{1})$ and $(\mathbf{k}_{2},\alpha _{2})$ in the $1S_{u}$
rest frame. This is a very violent decay, and we cannot simply assume
that the $ {}1S_{p}$ is created at rest. \ There could be significant
recoil and non-collinear alignment of the two photons. In the
c.m. system of the initial positronium atom $E_{{}1S_{p}}=\sqrt{\left(
  \sqrt{2}m\right) ^{2}+ \mathbf{p}_{{}1S_{p}}^{2}}$ and performing
the $d^{3}p_{{}1S_{p}}$ integral gives
\begin{equation}
d^{2}w=2\pi \lvert T_{fi}(\omega _{1},\omega _{2})\lvert
^{2}d^{3}k_{1}d^{3}k_{2}\delta (2m-E_{1S_{p}}-\omega _{1}-\omega
_{2})+O(\alpha ^{2}),
\end{equation}
in which\ 
\begin{align}
E_{1S_{p}}& =\sqrt{2m^{2}+\left( \mathbf{k}_{1}+\mathbf{k}_{2}\right) ^{2}}
+O(\alpha ^{2})  \notag \\
& =\sqrt{2m^{2}+\omega _{1}^{2}+\omega _{2}^{2}+2\omega _{1}\omega _{2}\cos
\theta _{12}}+O(\alpha ^{2}),  \notag \\
\cos \theta & =\cos \theta _{1}\cos \theta _{2}+\sin \theta _{1}\sin \theta
_{2}\cos (\phi _{1}-\phi _{2}).
\end{align}
With 
\begin{equation}
\int \int d^{3}k_{1}d^{3}k_{2}=\int_{0}^{\infty }d\omega
_{1}\int_{0}^{\infty }d\omega _{2}\omega _{1}^{2}\omega
_{2}^{2}\int_{0}^{\pi }\int_{0}^{\pi }\sin \theta _{1}\sin \theta
_{2}d\theta _{1}d\theta _{2}\int_{0}^{2\pi }\int_{0}^{2\pi }d\phi _{1}d\phi
_{2}.
\end{equation}
we perform the \ $\omega _{2}$ integral using
\begin{align}
\delta (2m-E_{{}1S_{p}}-\omega _{1}-\omega _{2})& =\delta (f(\omega _{2}))=
\frac{\delta (\omega _{2}-\omega _{20})}{\left\vert f~^{\prime }(\omega
_{2})\right\vert },  \notag \\
f(\omega _{2})& =2m-\sqrt{2m^{2}+\omega _{1}^{2}+\omega _{2}^{2}+2\omega
_{1}\omega _{2}\cos \theta _{12}}-\omega _{1}-\omega _{2},  \notag \\
\omega _{20}& =\frac{2m\omega _{1}-m^{2}}{(\omega _{1}(1-\cos \theta
_{12})-2m)},  \notag \\
f^{\prime }(\omega _{2})& =-\frac{2m-\omega _{1}(1-\cos \theta _{12})}{
\left( 2m-\omega _{1}-\omega _{2}\right) },
\end{align}
or
\begin{equation}
\left\vert \frac{1}{f^{\prime }(\omega _{20})}\right\vert =\left\vert \frac{
\left( \left( 2m-\omega _{1}\right) \omega _{1}(1-\cos \theta
_{12})-3m^{2}\right) }{\left( 2m-\omega _{1}(1-\cos \theta _{12})\right) ^{2}
}\right\vert \equiv g(\omega _{1}).
\end{equation}
Now, \ we must have $\omega _{20}={(2m\omega _{1}-m^{2})}/{[\omega
_{1}(1-\cos \theta _{12})-2m]}>0$. \ Thus we must have that either
\begin{align}
\omega _{1}& >\frac{m}{2}\equiv \omega _{0},  \notag \\
\text{and }\omega _{1}& >\frac{2m}{(1-\cos \theta _{12})}\equiv \tilde{\omega
}_{0},
\end{align}
or
\begin{align}
\omega _{1}& <\frac{m}{2}\equiv \omega _{0},  \notag \\
\text{and }\omega _{1}& <\frac{2m}{(1-\cos \theta _{12})}\equiv \tilde{\omega
}_{0}.
\end{align}
The first of these conditions is not possible because it would allow an $
\omega _{1}$ that is not bounded and would clearly not satisfy energy
conservation. \ Clearly $\omega _{0},\tilde{\omega}_{0}>0$. \ Let us compare 
$\omega _{0},\tilde{\omega}_{0}.$ We find $\omega _{0}<\tilde{\omega}_{0}$
would be true if 
\begin{equation}
\frac{4}{(1-\cos \theta _{12})}>1,
\end{equation}
which is true for all $\cos \theta _{12}$. Thus, $\omega _{0}<\tilde{\omega}
_{0}$ implies second set of inequalities are true if $\omega _{1}<\frac{m}{2}
.$ Thus we have
\begin{align}
& \int \int d^{3}k_{1}d^{3}k_{2}2\pi \lvert T_{fi}(\omega _{1},\omega
_{2})\lvert ^{2}\delta (2m-E_{1S_{p}}-\omega _{1}-\omega _{2}) \\
& =\int_{0}^{m/2}d\omega _{1}\omega _{1}^{2}g(\omega _{1})\int_{0}^{\pi
}\int_{0}^{\pi }\sin \theta _{1}\sin \theta _{2}d\theta _{1}d\theta _{2} 
\notag \\
& \times \int_{0}^{2\pi }\int_{0}^{2\pi }d\phi _{1}d\phi _{2}\left( \frac{
2m\omega _{1}-m^{2}}{(\omega _{1}(1-\cos \theta _{12})-2m)}\right) ^{2}2\pi
\lvert T_{fi}(\omega _{1},\frac{2m\omega _{1}-m^{2}}{(\omega _{1}(1-\cos
\theta _{12})-2m)})\lvert ^{2}.  \notag
\end{align}
Our desired matrix element is 
\begin{equation}
T_{fi}=\langle 1S_{p}\gamma \gamma \lvert H_{int}\lvert 1S_{u}\rangle
M_{\zeta \zeta ^{\prime }}.
\end{equation}
The interaction field theoretic Hamiltonian is 
\begin{align}
H_{int}& =-\frac{1}{4\mu }\left[ e\left( \mathbf{p\cdot A(x}_{1},t)+\mathbf{
A(x}_{1},t\mathbf{)\cdot p}\right) +e^{2}\mathbf{A}^{2}\mathbf{(x}_{1}
\mathbf{)}\right] \\
& -\frac{1}{4\mu }\left[ e\left( \mathbf{p\cdot A(x}_{2},t)+\mathbf{A(x}
_{2},t\mathbf{)\cdot p}\right) +e^{2}\mathbf{A}^{2}\mathbf{(x}_{2}\mathbf{)}
\right] ,  \notag
\end{align}
in which 
\begin{equation}
\mathbf{A}(\mathbf{x},t)=\frac{1}{(2\pi )^{3/2}}\sum_{\mathbf{k}
}\sum_{\alpha }\sqrt{\frac{1}{2\omega }}\big[a_{\mathbf{k},\alpha }(0)
\mathbf{\varepsilon }^{(\alpha )}e^{i\mathbf{k}\cdot \mathbf{x}-i\omega
t}+a_{\mathbf{k},\alpha }^{\dagger }(0)\mathbf{\varepsilon }^{(\alpha )}e^{-i
\mathbf{k}\cdot \mathbf{x}+i\omega t}\big].
\end{equation}

The only terms of $H_{int}$ that contribute are 
\begin{align}
H_{int} & =-\frac{e}{2\mu}\sum_{\mathbf{k},\alpha}\sqrt{\frac{1}{2\omega_{
\mathbf{k}}(2\pi)^{3}}}\{\frac{1}{2}\mathbf{\varepsilon}^{(\alpha )}\cdot
\left[ \mathbf{p~}a_{\mathbf{k},\alpha}^{\dagger}e^{-i\mathbf{k}\cdot\mathbf{
x}_{1}}+a_{\mathbf{k},\alpha}^{\dagger}e^{-i\mathbf{k}\cdot\mathbf{x}_{1}}
\mathbf{p}\right]  \notag \\
& +\frac{1}{2}\mathbf{\varepsilon}^{(\alpha)}\cdot\left[ \mathbf{p}~\mathbf{
\varepsilon}^{(\alpha)}a_{\mathbf{k},\alpha}^{\dagger}e^{-i\mathbf{k}\cdot
\mathbf{x}_{2}}+a_{\mathbf{k},\alpha}^{\dagger}e^{-i\mathbf{k}\cdot\mathbf{x}
_{2}}\mathbf{p}\right] \}  \notag \\
& +\frac{e^{2}}{m}\sum_{\mathbf{k},\alpha}\sum_{\mathbf{k}
^{\prime},\alpha^{\prime}}\frac{1}{2\sqrt{\omega_{\mathbf{k}}\omega_{\mathbf{
k}^{\prime}}}(2\pi)^{3}}\mathbf{\varepsilon}^{(\alpha)}\cdot\mathbf{
\varepsilon }^{(\alpha^{\prime})}  \notag \\
& \times\{\frac{1}{2}\bigg(a_{\mathbf{k},\alpha}^{\dagger}a_{\mathbf{k}
^{\prime},\alpha^{\prime}}^{\dagger}e^{-i(\mathbf{k}+\mathbf{k}
^{\prime})\cdot\mathbf{x}_{1}}\bigg)+\frac{1}{2}\bigg(a_{\mathbf{k}
,\alpha}^{\dagger }a_{\mathbf{k}^{\prime},\alpha^{\prime}}^{\dagger}e^{-i(
\mathbf{k}+\mathbf{k}^{\prime})\cdot\mathbf{x}_{2}}\bigg)\}  \notag \\
& =H_{I1}+H_{I2}.
\end{align}
$H_{I1}$ creates only one photon in first order, so we must use second order
perturbation for this term. $H_{I2}$ creates two photons, so we need only
consider its first order contribution. 
\begin{align}
T_{fi} & =M_{\zeta\zeta'}\langle1S_{p}\lvert H_{I2}\lvert1S_{u}\rangle+M_{\zeta\zeta'}\sum_{I}\frac{
\langle1S_{p}\lvert H_{I1}\lvert I\rangle\langle I\lvert
H_{I1}\lvert1S_{u}\rangle}{E_{1S_{u}}-E_{I}}  \notag \\
& =\bigg(\frac{e^{2}}{m}\bigg)\bigg(\frac{M_{\zeta\zeta'}}{2(2\pi)^{3}\sqrt{\omega
_{1}\omega_{2}}}\bigg)\bigg[\mathbf{\varepsilon}^{(\alpha_{1})}\cdot \mathbf{
\varepsilon}^{(\alpha_{2})}\langle1S_{p}\lvert\frac{1}{2}e^{-i(\mathbf{k}
_{1}+\mathbf{k}_{2})\cdot\mathbf{x}_{1}}+\frac{1}{2}e^{-i(\mathbf{k}_{1}+
\mathbf{k}_{2})\cdot\mathbf{x}_{2}}\lvert1S_{u}\rangle  \notag \\
& +\frac{M_{\zeta\zeta'}}{m}\sum_{I}\frac{1}{E_{1S_{u}}-E_{I}-\hbar\omega_{1}}\bigg(
\langle1S_{p}\lvert\mathbf{\varepsilon}^{(\alpha_{2})}\cdot(\frac{1}{2}\left[
\mathbf{p~}e^{-i\mathbf{k}_{2}\cdot\mathbf{x}_{1}}+e^{-i\mathbf{k}_{2}\cdot
\mathbf{x}_{1}}\mathbf{p}\right] +\frac{1}{2}\left[ \mathbf{p~}e^{-i\mathbf{k
}_{2}\cdot\mathbf{x}_{2}}+e^{-i\mathbf{k}_{2}\cdot\mathbf{x}_{2}}\mathbf{p}
\right] )\lvert I\rangle  \notag \\
& \times\langle I\lvert\mathbf{\varepsilon}^{(\alpha_{1})}\cdot(\frac{1}{2}
\left[ \mathbf{p~}e^{-i\mathbf{k}_{1}\cdot\mathbf{x}_{1}}+e^{-i\mathbf{k}
_{1}\cdot\mathbf{x}_{1}}\mathbf{p}\right] +\frac{1}{2}\left[ \mathbf{p~}e^{-i
\mathbf{k}_{1}\cdot\mathbf{x}_{2}}+e^{-i\mathbf{k}_{1}\cdot\mathbf{x}_{2}}
\mathbf{p}\right] )\lvert1S_{u}\rangle\bigg)  \notag \\
& +\frac{M_{\zeta\zeta'}}{m}\sum_{I}\frac{1}{E_{1S_{u}}-E_{I}-\hbar\omega_{2}}\bigg(
\langle1S_{p}\lvert\mathbf{\varepsilon}^{(\alpha_{1})}\cdot(\frac{1}{2}\left[
\mathbf{p~}e^{-i\mathbf{k}_{1}\cdot\mathbf{x}_{1}}+e^{-i\mathbf{k}_{1}\cdot
\mathbf{x}_{1}}\mathbf{p}\right] +\frac{1}{2}\left[ \mathbf{p~}e^{-i\mathbf{k
}_{1}\cdot\mathbf{x}_{2}}+e^{-i\mathbf{k}_{2}\cdot\mathbf{x}_{2}}\mathbf{p}
\right] )\lvert I\rangle  \notag \\
& \times\langle I\lvert\mathbf{\varepsilon}^{(\alpha_{2})}\cdot(\frac{1}{2}
\left[ \mathbf{p~}e^{-i\mathbf{k}_{2}\cdot\mathbf{x}_{1}}+e^{-i\mathbf{k}
_{2}\cdot\mathbf{x}_{1}}\mathbf{p}\right] +\frac{1}{2}\left[ \mathbf{p~}e^{-i
\mathbf{k}_{2}\cdot\mathbf{x}_{2}}+e^{-i\mathbf{k}_{2}\cdot\mathbf{x}_{2}}
\mathbf{p}\right] )\lvert1S_{u}\rangle\bigg)\bigg),
\end{align}

\noindent in which
\begin{equation}
E=b^{2}/2\mu =b^{2}/m.
\end{equation}
In creating two photons, the second order contributions includes the
emission of $(\mathbf{k}_{1},\alpha _{1})$ followed by the emission of
$( \mathbf{k}_{2},\alpha _{2})$ and also the emission of $(\mathbf{k}
_{2},\alpha _{2})$ followed by the emission of $(\mathbf{k}_{1},\alpha
_{1})$ . \ We use the dipole approximation\footnote{ We compute $k\sim
  m(1-\sqrt{2}/2),$ and with $\psi _{1S_{p}}\sim \exp (-mr),$ and so
  $kr\sim (1-\sqrt{2}/2)\sim 0.3$ and justification for the dipole
  approximation is not as clear cut as with the meta-stable decay of
  the usual state where $k\sim m\alpha ^{2},\psi _{1S_{u}}\sim \exp
  (-m\alpha r),$ $ kr\sim \alpha $ but for our purposes it is
  sufficient.},
\begin{equation}
e^{-i\mathbf{k}_{i}\cdot \mathbf{x}_{i}}\simeq 1.
\end{equation}
In that case in terms of the matrix elements of the relative coordinate $
\mathbf{x}$, we have 
\begin{align}
\langle {}1S_{p}\lvert \mathbf{p}\cdot \mathbf{\varepsilon }^{(\alpha
_{2})}\lvert I\rangle & =-\mu i\langle {}1S_{p}\lvert \lbrack \mathbf{r,}
H_{0}]\cdot \mathbf{\varepsilon }^{(\alpha _{2})}\lvert I\rangle =\frac{1}{2}
im(E_{{}1S_{p}}-E_{I})\langle {}1S_{p}\lvert \mathbf{x}\lvert I\rangle \cdot 
\mathbf{\varepsilon }^{(\alpha _{2})},  \notag \\
\langle I\lvert \mathbf{p}\cdot \mathbf{\varepsilon }^{(\alpha _{1})}\lvert
{}1S_{u}\rangle & =\frac{1}{2}im(E_{I}-E_{{}1S_{u}})\langle I\lvert \mathbf{x
}\lvert {}1S_{u}\rangle \cdot \mathbf{\varepsilon }^{(\alpha _{1})},
\end{align}
so that 
\begin{align}
T_{fi}& \simeq \bigg(\frac{e^{2}}{m}\frac{1}{2(2\pi )^{3}\sqrt{\omega _{1}\omega
_{2}}}\bigg)M_{\zeta \zeta ^{\prime }}
\bigg[\mathbf{\varepsilon }^{(\alpha _{1})}\cdot \mathbf{
\varepsilon }^{(\alpha _{2})}\langle {}1S_{p}\lvert 1S_{u}\rangle  \notag \\
& -m\sum_{I}(E_{{}1S_{p}}-E_{I})(E_{I}-E_{{}^{1}{}1S_{u}})\bigg(\frac{
\langle 1S_{p}\lvert \mathbf{x}\lvert I\rangle \cdot \mathbf{\varepsilon }
^{(\alpha _{2})}\langle I\lvert \mathbf{x}\lvert {}1S_{u}\rangle \cdot 
\mathbf{\varepsilon }^{(\alpha _{1})}}{E_{{}1S_{u}}-E_{I}-\hbar \omega _{1}}
\notag \\
& +\frac{\langle {}1S_{p}\lvert \mathbf{x}\lvert I\rangle \cdot \mathbf{
\varepsilon }^{(\alpha _{1})}\langle I\lvert \mathbf{x}\lvert
{}1S_{u}\rangle }{E_{{}1S_{u}}-E_{I}-\hbar \omega _{2}}\bigg)\bigg].
\end{align}

The allowed dipole transitions are ($l_{f}=l_{i}\pm
1$;~$m_{f}=m_{i},m_{i} \pm 1$), so from parity considerations only
$l=1,3,..$ intermediate states yield non-vanishing contributions. The
decay from ${}1S_{u}$ to ${}1S_{p}$ consists of a direct term and a
combined transition, first from the $ {}1S_{u} $ state to a virtual
$P$ state, then from the $P$ state to the peculiar ground state.

To get an estimate of the relative size of the direct and virtual
contributions we consider only the lowest lying \ $P$ state and
neglect the polarization factors and approximate $\mathbf{r\rightarrow
}z\mathbf{\hat{k}} $.  We thus approximate the amplitude by
\begin{align}
T_{fi}& \simeq \bigg(\frac{e^{2}}{m}\frac{1}{2(2\pi )^{3}\sqrt{\omega
_{1}\omega _{2}}}\bigg) M_{\zeta \zeta ^{\prime }} \bigg[\langle 1S_{p}\lvert
1S_{u}\rangle - m(E_{{}1S_{p}}-E_{2P})(E_{2P}-E_{{}1S_{u}})  \notag \\
& \times \bigg(\frac{\langle {}1S_{p}\lvert z\lvert 2P\rangle \langle
2P\lvert z\lvert 1S_{u}\rangle }{E_{{}1S_{u}}-E_{{}2P}-\hbar \omega _{1}}+
\frac{\langle {}1S_{p}\lvert z\lvert 2P\rangle \langle {}2P\lvert z\lvert
{}1S_{u}\rangle }{E_{{}1S_{u}}-E_{{}2P}-\hbar \omega _{2}}\bigg)\bigg].
\end{align}

Given \cite{pec} 
\begin{align}
\psi _{1S_{u}}& =\frac{1}{\sqrt{4\pi }}\left[ \left( \frac{2\varepsilon
_{w_{+}}\alpha }{n_{+}^{\prime }}\right) ^{3}\frac{n_{r}!}{2n_{+}^{\prime
}(n_{+}^{\prime }+\lambda _{+})!}\right] ^{1/2}\exp (-\frac{\varepsilon
_{w_{+}}\alpha r}{n_{+}^{\prime }})\left( \frac{2\varepsilon _{w_{+}}\alpha r
}{n_{+}^{\prime }}\right) ^{\lambda _{\pm }},  \notag \\
n_{+}^{\prime }& =(1+\sqrt{1-4\alpha ^{2}})/2\sim 1,~\varepsilon
_{w_{+}}\sim m/2;  \notag \\
~\left( n_{+}^{\prime }+\lambda _{+}\right) !& =\left( 2\lambda
_{+}+2-1\right) !=(\sqrt{1-4\alpha ^{2}})!\sim 1,  \notag \\
\psi _{1S_{p}}& =\frac{1}{\sqrt{4\pi }}\left[ \left( \frac{2\varepsilon
_{w_{-}}\alpha }{n_{-}^{\prime }}\right) ^{3}\frac{n_{r}!}{2n_{-}^{\prime
}(n_{-}^{\prime }+\lambda _{-})!}\right] ^{1/2}\exp (-\frac{\varepsilon
_{w_{-}}\alpha r}{n_{-}^{\prime }})\left( \frac{2\varepsilon _{w_{-}}\alpha r
}{n_{-}^{\prime }}\right) ^{\lambda _{-}}  \notag \\
n_{-}^{\prime }& =(1-\sqrt{1-4\alpha ^{2}})/2\sim \alpha ^{2},~\varepsilon
_{w_{-}}\sim m\alpha /\sqrt{2};~  \notag \\
\left( n_{-}^{\prime }+\lambda _{-}\right) !& =\left( 2\lambda
_{-}+2-1\right) !=(-\sqrt{1-4\alpha ^{2}})!\sim (-1+2\alpha ^{2})!=\frac{
(2\alpha ^{2})!}{2\alpha ^{2}}\sim \frac{1}{2\alpha ^{2}},  \label{upsn}
\end{align}
so that
\begin{align}
\psi _{1S_{u}}& \sim \frac{1}{\sqrt{\pi }}\frac{(m\alpha )^{3/2}}{2\sqrt{2}}
\exp (-\frac{m\alpha r}{2}),  \notag \\
\psi _{1S_{p}}& \sim \frac{1}{\sqrt{4\pi }}\left[ \left( \sqrt{2}m\right)
^{3}\right] ^{1/2}\exp (-\frac{mr}{\sqrt{2}})\left( \sqrt{2}mr\right)
^{(-1+\alpha ^{2})},  \label{upfn}
\end{align}
and hence $\ $with $(\frac{1}{\sqrt{2}}+\alpha /2)mr=x$, we have 
\begin{align}
\langle 1S_{p}\lvert 1S_{u}\rangle & \sim \frac{(m\alpha )^{3/2}}{4\pi \sqrt{
2}}\left( \sqrt{2}m\right) ^{3/2}\int d^{3}r\exp (-(\frac{1}{\sqrt{2}}
+\alpha /2)mr)\left( \sqrt{2}mr\right) ^{(-1+\alpha ^{2})}  \notag \\
& \sim \frac{2^{1/4}\alpha ^{3/2}}{4\pi }2\sqrt{2}\times 4\pi \times \frac{1
}{\sqrt{2}}\int xdx\exp (-x)=2^{5/4}\alpha ^{3/2}.
\end{align}
Next consider the matrix elements\ $\langle 1S_{p}\lvert z\lvert
2P(m=0)\rangle $ and $\langle 2P(m=0)\lvert z\lvert 1S_{u}\rangle .~$The
second one involves only usual states and roughly just one size (the Bohr
radius). \ The $2P(m=0)$ state is
\begin{equation*}
\psi _{2P0}(\mathbf{r)}\mathbf{=}\frac{1}{4\sqrt{2\pi }}\left( \frac{m\alpha 
}{2}\right) ^{3/2}\frac{rm\alpha }{2}\exp (-rm\alpha /4)\cos \theta .
\end{equation*}
Thus 
\begin{align}
\langle 2P(m& =0)\lvert r\cos \theta \lvert 1S_{u}\rangle =\frac{1}{4\sqrt{
2\pi }}\left( \frac{m\alpha }{2}\right) ^{3/2}\frac{1}{\sqrt{\pi }}\frac{
(m\alpha )^{3/2}}{2\sqrt{2}}  \notag \\
& \times \int d^{3}r\exp (-\frac{m\alpha r}{2})\frac{rm\alpha }{2}\exp
(-rm\alpha /4)r\cos ^{2}\theta  \notag \\
& =\frac{1}{\sqrt{2}}\frac{64}{729m\alpha }\int_{0}^{\infty }dxx^{4}\exp
(-x)\sim \frac{1.5}{m\alpha }.
\end{align}
Using $({1}/\sqrt{2}+\alpha /4)mr=x$, the second one is 
\begin{align}
\langle 1S_{p}\lvert z|2P\rangle & =\frac{1}{8\sqrt{2\pi }}\left( \frac{
m\alpha }{2}\right) ^{3/2}\frac{1}{\sqrt{\pi }}\left( \sqrt{2}m\right)
^{3/2}\int d^{3}rr\exp (-\frac{mr}{\sqrt{2}})\left( \sqrt{2}mr\right)
^{(-1+\alpha ^{2})}  \notag \\
& \times \frac{rm\alpha }{2}\exp (-rm\alpha /4)\cos ^{2}\theta  \notag \\
& =\frac{(m^{3}\alpha ^{3/2})\alpha }{2^{3/4}24\left[ (\frac{1}{\sqrt{2}}
+\alpha /4)m\right] ^{4}}\int dxx^{3}\exp (-x)\sim \frac{\alpha ^{5/2}}{
2^{3/4}m}.
\end{align}
Combining all factors we have in terms of orders of $\alpha $
\begin{align}
T_{fi}\simeq \bigg(\frac{e^{2}}{m}\frac{1}{2(2\pi )^{3}\sqrt{\omega
_{1}\omega _{2}}}\bigg)M_{\zeta \zeta ^{\prime }} \bigg[2^{5/4}\alpha ^{3/2}&
-m(E_{1S_{p}}-E_{2P})(E_{2P}-E_{1S_{u}}) 
\notag \\
& \times \frac{1.5}{m\alpha }\frac{\alpha ^{5/2}}{2^{3/4}m}\bigg(\frac{1}{
E_{1S_{u}}-E_{2P}-\hbar \omega _{1}}+\frac{1}{E_{1S_{u}}-E_{2P}-\hbar \omega
_{2}}\bigg)\bigg].  \notag \\
&
\end{align}
From
\begin{align}
E_{2P}-E_{1S_{u}}& \sim m\alpha ^{2},  \notag \\
E_{1S_{p}}-E_{2P}& \sim m\left( 2-\sqrt{2}\right) ,
\end{align}
it is clear that the second terms are order $\alpha ^{2}$ smaller than the
first. We ignore these higher order pieces. \ Thus we approximate
\begin{equation}
T_{fi}\simeq \bigg(\frac{e^{2}}{m}\frac{1}{2(2\pi )^{3}\sqrt{\omega
_{1}\omega _{2}}}\bigg)M_{\zeta \zeta ^{\prime }} \bigg[2^{5/4}\alpha ^{3/2}
\bigg].
\end{equation}
With this we compute 
\begin{align}
\Gamma & =\int \int d^{3}k_{1}d^{3}k_{2}2\pi \lvert T_{fi}(\omega
_{1},\omega _{2})\lvert ^{2}\delta (2m-E_{1S_{p}}-\omega _{1}-\omega
_{2})P_{up}  \notag \\
& =\int_{0}^{m/2}d\omega _{1}\omega _{1}g(\omega
_{1}))\int_{-1}^{+1}dz\left( \frac{-m^{2}+2m\omega _{1}}{(\omega
_{1}(1-z)-2m)}\right) \frac{16\pi ^{2}\alpha ^{2}}{m^{2}}\frac{\alpha ^{3}
\sqrt{2}}{4\pi ^{3}}P_{up}.
\end{align}

\bigskip We change the radial integration variable to $x=\omega _{1}/m$ so
that, with 
\begin{align}
g(\omega _{1})& =\left\vert \frac{\left( \left( 2m-\omega _{1}\right) \omega
_{1}(1-\cos \theta _{12})-3m^{2}\right) }{\left( 2m-\omega _{1}(1-\cos
\theta _{12})\right) ^{2}}\right\vert  \notag \\
& =\left\vert \frac{\left( \left( 2-x\right) x(1-z)-3\right) }{\left(
2-x(1-z)\right) ^{2}}\right\vert ,
\end{align}
we have
\begin{equation}
\Gamma =\frac{4m\alpha ^{5}\sqrt{2}}{\pi }\int_{0}^{1/2}dxx\int_{-1}^{+1}dz
\left( \frac{2x-1}{(x(1-z)-2)}\right) \left\vert \frac{\left( \left(
2-x\right) x(1-z)-3\right) }{\left( 2-x(1-z)\right) ^{2}}\right\vert P_{up}.
\end{equation}
This requires the integral 
\begin{equation}
\int_{0}^{1/2}dxx\int_{-1}^{+1}dz\left( \frac{2x-1}{(x(1-z)-2)}\right)
\left\vert \frac{\left( \left( 2-x\right) x(1-z)-3\right) }{\left(
2-x(1-z)\right) ^{2}}\right\vert \sim 4.17\times 10^{-2}.
\end{equation}
The decay rate is thus 
\begin{align}
\Gamma & =\frac{4m\alpha ^{5}\sqrt{2}}{\pi }4.17\times 10^{-2}P_{up}  \notag
\\
& =\Gamma _{1S_{u}~\rightarrow ~2\gamma \times }0.152~P_{up}
\end{align}
and so the branching ratio is $0.152P_{up}$.

\bigskip

\end{document}